\documentclass[10pt,conference]{IEEEtran} 
\usepackage{graphicx}
\usepackage{balance} 
\usepackage{amssymb}
\usepackage{amscd}
\usepackage{dsfont}
\usepackage{amsmath}
\usepackage{epsfig}
\usepackage{graphics}
\usepackage{psfrag}
\usepackage{rotating}
\usepackage{amsmath}
\usepackage{amsfonts}
\usepackage{url}
\usepackage{color}

\newtheorem{theorem}{Theorem}
\newtheorem{definition}[theorem]{Definition}

\newtheorem{corollary}[theorem]{Corollary}

\newcommand{\bs}{\boldsymbol}
\newcommand{\mc}{\mathcal}

\newcommand{\defn}{\stackrel{\triangle}{=} }
\newcommand{\ds}{\displaystyle}
\newcommand{\BR}{\mathrm{BR}}

\newcommand{\diag}{\mathrm{diag}}

\newcommand{\gameone}{\mathcal{G}^{(a)}}
\newcommand{\gametwo}{\mathcal{G}^{(b)}}
\newcommand{\gamei}{\mathcal{G}^{(i)}}

\begin{document}
\title{On the Nash Equilibria in Decentralized Parallel Interference Channels}

\author{
\IEEEauthorblockN{Luca Rose}
\IEEEauthorblockA{Alcatel - Lucent Chair in Flexible Radio\\
Supelec, France.\\
luca.rose@supelec.fr}
\and
\IEEEauthorblockN{Samir~M. Perlaza}
\IEEEauthorblockA{Orange Labs - Paris,\\
France Telecom R\&D, France.\\
samir.medinaperlaza@orange-ftgroup.com}
\and
\IEEEauthorblockN{M\'{e}rouane Debbah}
\IEEEauthorblockA{Alcatel - Lucent Chair in Flexible Radio\\
Supelec, France.\\
merouane.debbah@supelec.fr}
}
\maketitle

\begin{abstract}
\boldmath
In this paper, the $2$-dimensional decentralized parallel interference channel (IC) with $2$ transmitter-receiver pairs is modelled as a non-cooperative static game. Each transmitter is assumed to be a fully rational entity with complete information on the game, aiming to maximize its own individual spectral efficiency by tuning its own power allocation (PA) vector. Two scenarios are analysed. First, we consider that transmitters can split their transmit power between both dimensions (PA game). Second, we consider that each transmitter is limited to use only one dimension (channel selection CS game). In the first scenario, the game might have either one or three NE in pure strategies (PS). However, two or infinitely many NE in PS might also be observed with zero probability. In the second scenario, there always exists either one or two NE in PS. We show that in both games there always exists a non-zero probability of observing more than one NE. More interestingly, using Monte-Carlo simulations, we show that the highest and lowest network spectral efficiency at any of the NE in the CS game are always higher than the ones in the PA.\\
Keywords: Interference Channel, Decentralized Network, Nash Equilibrium, Braess Paradox, Spectrum Efficiency.
\end{abstract}

\section{Introduction}\label{SecIntroduction}
This article addresses the interaction of $2$ transmitters-receiver links subject to mutual interference due to the use of $2$ common frequency bands. Here, each transmitter communicates only with its corresponding receiver aiming to maximize its individual spectral efficiency (ISE) regardless of the ISE achieved by the other link. We do not consider neither transmitter nor receiver cooperation and any kind of message exchanging between transmitters is completely avoided. Thus, this interaction through mutual interference is modelled by two different static strategic games. In the first game, we consider that transmitters can simultaneously use both channels (frequency bands). In the second game, we limit each transmitter to use only one of the two available channels. In the following, we refer to the former as the power allocation (PA) game and the latter as channel selection (CS) game. In both games, transmitters are assumed to be fully rational entities with complete information. This assumption might appear not practically appealing, however, our interest focuses in identifying the set of Nash equilibria of each of the games in order to compare the system spectral efficiency, i.e., the sum of all the ISE, achieved in each game. Any mechanism or algorithm for achieving NE under real-system implementation constraints is considered out of the scope of this paper. The interested reader is referred to \cite{Scutari-Algorithms-2008, Scutari-Palomar-09} and references therein.
\noindent
Regarding the existence and uniqueness of the NE in the PA game, many results are already known for more general cases than the one described here. In the most general case, i.e., the $K$-transmitter MIMO decentralized interference channel, the existence of the NE has been already proved in \cite{Scutari-Palomar-09}. The same holds for the interference relay channel described in \cite{belmega-gamenets-2009}. However, regarding the multiplicity of the NE in the PA game, much less is known. For instance, in \cite{Scutari-Palomar-09} sufficient but not necessary conditions for the uniqueness are provided and, aside from the result in \cite{belmega-gamenets-2009}, the exact number of NE in pure strategies, in a general context, remains an open problem.

\noindent
In the CS game, conversely to the PA one, no results are known with respect to the existence or multiplicity of the NE. The relevance of the CS game relies on the fact that under non-perfect channel estimations, the water-filling PA \cite{Scutari-Palomar-09} can not be implemented, and thus, transmitters must either transmit over a single channel e.g., Wi-Fi networks, or to use predefined power allocation vectors. More importantly, it has been shown that significant benefits from the global system point of view are obtained by limiting the transmitters to use a reduced number of channels, at least in the parallel multiple access channel \cite{Perlaza-Crowncom-09, Perlaza-Gamecomm-09}. This result implies the existence of a Braess type paradox \cite{Braess-69}, since reducing the set of actions of each player leads to a better global performance. The existence of this paradox has been already reported for specific channel realizations in the interference channel \cite{Altman-wiopt-2008, Altman-wiopt-08b}.    

\noindent
The main contributions of this paper can be listed as follows. $(i)$ Contrary to previous beliefs \cite{bennis-gamenet-2009}, the number of NE in the PA game is shown to be $1$ or $3$ depending on the exact channel realizations. However, with zero-probability, it is also possible to observe either $2$ or infinitely many NE. This result aligns with the number of NE in the interference relay channel described in \cite{belmega-gamenets-2009}. $(ii)$ Depending on the channel realizations, any feasible channel selection in the CS game might be a NE. Here, we provide conditions over the channel realizations for every case.  Moreover for any channel realization, it always exist at least one NE in pure strategies. $(iii)$ The number of NE in the CS game is either one or two depending on the channel realizations. $(iv)$ The best and worst average system spectral efficiency achieved in equilibrium in the CS game is better than the best and worst average system spectral efficiency achieved in equilibrium in the PA game.  

\noindent
The paper is organized as follows. In Sec. \ref{SecSystemModel}, we describe the decentralized parallel interference channel addressed in this paper. In Sec. \ref{SecGameFormulation}, we present the formal game formulation of both the PA game and CS game. In Sec. \ref{SecGa} and \ref{SecGb}, we provide the main results regarding the existence and multiplicity of the NE in the PA game and CS game, respectively. In \ref{SecSimulations}, we use Monte-Carlo methods to identify the probability of the different number of NE in both the PA and CS games. We also compare using Monte-Carlo simulations the system spectral efficiency achieved by both games. This paper is concluded by Sec. \ref{SecConclusions}.

\section{System model}\label{SecSystemModel}

Consider a set $\mathcal{K} \defn \lbrace 1, 2 \rbrace$ of transmitter-receiver pairs. Each transmitter sends private information to its respective receiver throughout a set $\mathcal{S} \defn \lbrace 1, 2 \rbrace$ of orthogonal channels. Here, the channel orthogonality is assumed in the frequency domain and transmissions take place simultaneously, thus communications are subject to mutual interference. Denote by $\bs{y}_j = \left( y_{j}^{(1)}, y_{j}^{(2)}\right)^T$ the $2$-dimensional vector representing the received signal at receiver $j \in \mathcal{K}$. Hence, $\bs{y}_{j}$ can be written in the baseband at the symbol rate as follows,
\begin{equation}
\bs{y}_j = \displaystyle\sum_{k = 1}^2 \bs{H}_{j,k} \bs{x}_k + \bs{z}_j.
\end{equation}
Here, $\forall (j,k) \in \mathcal{K}^2$, the matrix $\bs{H}_{j,k}$ is the channel transfer matrix from transmitter $k$ to the receiver $j$, and $\bs{H}_{j,k} = \diag\left(h_{j,k}^{(1)}, h_{j,k}^{(2)}\right)$. Besides $\forall (j,k,s) \in \mathcal{K}^2 \times\mathcal{S}$, $h_{j,k}^{(s)}$ represents the channel realization between transmitter $k$ and receiver $j$ over channel $s$. In our analysis, flat fading channels are assumed, i.e., each channel realization is time-invariant over the whole channel use (e.g., frame length). The entries $h_{j,k}^{(s)}$ are time-invariant realizations of a complex circularly symmetric Gaussian random variable, with zero mean and unit variance. 
The vector $\bs{x}_k = \left( x_{k}^{(1)},  x_{k}^{(2)}\right)$ is the vector of symbols transmitted by transmitter $k$. For all $s \in \mathcal{S}$, $x_{k}^{(s)}$ represents the symbol sent by transmitter $k$ over channel $s$. Here, 
$\bs{x}_k$ is a $2$-dimensional complex circularly symmetric Gaussian random variable with zero mean and covariance matrix $\bs{P}_k = \mathds{E}\left(\bs{x}_k \bs{x}_k^*\right) = \diag\left( p_{k}^{(1)}, p_{k}^{(2)}\right)$.  For all $(k,s) \in \mathcal{K}\times\mathcal{S}$, $p_{k}^{(s)}$ represents the transmit power allocated by transmitter $k$ over channel $s$. Transmitters are power-limited, that is,
\begin{equation}
\forall k \in \mathcal{K}, \quad p_{k}^{(1)} + p_{k}^{(2)} \leqslant p_{k,\max},
\label{EqPowerConstraints}
\end{equation}
where $p_{k,\max}$ is the maximum transmit power of transmitter $k$.
A power allocation (PA) vector for transmitter $k \in \mathcal{K}$ is any vector $\bs{p}_k = \left(p_{k}^{(1)}, p_{k}^{(2)}\right)$ with non-negative entries satisfying \eqref{EqPowerConstraints}.
The noise vector $\bs{z}_j$, $j\in\mathcal{K}$, is a $2$-dimensional zero mean Gaussian random variable with independent, equal variance real and imaginary parts. Here, $\mathds{E}\left(\bs{z}_j \bs{z}_j^* \right) = \diag\left(({\sigma^{(1)}_{j}})^2,({\sigma^{(2)}_{j}})^2\right)$, where, $({\sigma^{(s)}_j})^2$ represents the noise power in the receiver $j$ over channel $s$.

We denote the individual spectral efficiency (in [bps/Hz]) of each transmitter $k$ as follows:
\begin{equation}\label{EqUtility}
u_k(\bs{p}_{k},\bs{p}_{-k}) = \displaystyle\sum_{s \in \mathcal{S}}
\log_2\left(1 + \frac{p_{k}^{(s)} \left| h_{k,k}^{(s)}\right|^2}{({\sigma_{k}^{(s)}})^2 + p_{-k}^{(s)} \left| h_{k,-k}^{(s)}\right|^2}\right).
\end{equation}
\noindent
In the following, we focus on the scenario where each transmitter $k\in\mathcal{K}$ aims at maximizing its individual spectral efficiency (\ref{EqUtility}) by tuning its corresponding power policy. We consider two problems:\\
$(i)$ \textbf{The Power Allocation (PA) Problem:} where, each transmitter $k$ is allowed to tune its power allocation vector $\bs{p}_k$ splitting the maximum available power into both channels,  i.e., ${p}_k^{(s)} \in[0,p_{k,max}]$;\\
$(ii)$ \textbf{The Channel Selection (CS) Problem:} where, each transmitter $k$ is limited into using only one channel at a time with full power, i.e., ${p}_k^{(s)} \in \left\lbrace 0,p_{k,max}\right\rbrace$.


\section{Normal-Form Game Formulation}\label{SecGameFormulation}

The PA and CS problems described in Sec. \ref{SecSystemModel} can be
respectively modelled by the following two non-cooperative static
games in strategic form (with $i \in \{a,b\}$):
\begin{equation}
\mc{G}^{(i)} = \left(\mathcal{K},
\left(\mathcal{P}^{(i)}_k\right)_{k \in \mathcal{K}},
\left(u_k\right)_{k \in \mathcal{K}}\right).
\end{equation}
In both games, the set of transmitters $\mathcal{K}$ is the set of players. An action of a given transmitter $k$ is a particular PA scheme, i.e., a $2$-dimensional PA vector $\bs{p}_k = \left(p_{k}^{(1)}, p_{k}^{(2)}\right) \in \mathcal{P}_k^{(i)}$, where $\mathcal{P}_k^{(i)}$ is the set of all possible PA vectors which transmitter $k$ can use either in the game $\gameone$ ($i = a$) or in the game $\gametwo$ ($i =b$). An action profile of the game $i \in \lbrace a, b \rbrace$ is a super vector
$$\bs{p} =\left(\bs{p}_{1}, \bs{p}_{2}\right) \in \mathcal{P}^{(i)},$$
where $\mathcal{P}^{(i)}$ is a  set obtained from the Cartesian product of the action sets, i.e., $\mathcal{P}^{(i)} = \mathcal{P}_1^{(i)} \times \mathcal{P}_{2}^{(i)}$.


The utility function for player $k$ in the games $\gamei$ is denoted by $u_k: \mathcal{P}^{(i)} \rightarrow \mathds{R}$, with $i \in \lbrace 1, 2\rbrace$ and corresponds to the individual spectral efficiency of transmitter $k$ (\ref{EqUtility}).

\noindent 
The solution concept used in this paper is that of Nash equilibrium (NE) \cite{Nash-1950}. A NE is an action profile $\bs{p} \in \mathcal{P}^{(i)}$, $i \in \lbrace 1,2 \rbrace$, such that, no player would increase its individual utility by unilateral deviation.
\begin{definition}[Pure Nash Equilibrium]\label{DefNE} \emph{In the
non-cooperative games in strategic form $\mathcal{G}^{(i)}$, with $i
\in \lbrace a, b \rbrace$, an action profile $\boldsymbol{p} \in
\mathcal{P}^{(i)}$ is an NE if it satisfies, for all $k \in
\mathcal{K}$ and for all $\bs{p}'_k \in \mathcal{P}_k^{(i)}$, that
\begin{equation}
u_k(\bs{p}_k,\bs{p}_{-k}) \geqslant  u_k(\bs{p}'_k,\bs{p}_{-k}).
\end{equation}
}
\end{definition}

\noindent
Note that, from Def. \ref{DefNE}, it becomes clear that, at the NE, each player's action is the best response to the actions taken by all the other players. 
An alternative definition of the NE can be stated using the concept of best response correspondence, which we define as follows,
\begin{definition}[Best-Response Correspondence]\label{DefBR} \emph{In the
non-cooperative games in strategic form $\mathcal{G}^{(i)}$, with $i
\in \lbrace a, b \rbrace$, the relation $\BR^{(i)}_k: \mathcal{P}_{-k}^{(i)} \rightarrow \mathcal{P}_{k}^{(i)}$ such that
    \begin{equation}\label{EqBRi}
        \BR^{(i)}_{k}\left(\bs{p}_{-k}\right) =
         \ds\arg\max_{\bs{q}_k \in \mathcal{P}^{(i)}_k}
            u_k\left(\bs{q}_k,\bs{p}_{-k}\right),
    \end{equation}
is defined as the best-response correspondence of player $k \in
\mathcal{K}$, given the actions $\bs{p}_{-k}\in\mathcal{P}_{-k}^{(i)}$ adopted by all the other players. 
}
\end{definition}
Note that we denote by $-k$ the user other than $k$. For all $i \in \lbrace a, b \rbrace$, let the action profile $\bs{p}^* \in \mathcal{P}^{(i)}$ be an NE and let the correspondence $\BR: \mathcal{P}^{(i)} \rightarrow \mathcal{P}^{(i)}$ be defined by 
$\BR\left(\bs{p}\right) = \lbrace \bs{q} \in \mathcal{P}^{(i)}:q_1 \in \BR_1^{(i)}(p_2)$ and $q_2 \in \BR_2^{(i)}(p_1)\rbrace$.
Hence, $\forall k \in \mathcal{K}$, it holds that
\begin{equation}\label{EqFixPointNE}
	\bs{p}^* \in \BR\left(\bs{p}^*\right).
\end{equation}

Finally, in both games, the PA vector of player $k$, $\bs{p}_k' = \left( p_{k}'^{(1)}, p_{k}'^{(2)}\right)$, such that $p_{k}'^{(1)} + p_{k}'^{(2)} < p_{k,\max}$ is strictly dominated by a vector $\bs{p}_k = \left( p_{k}'^{(1)} + \epsilon_1 , p_{k}'^{(2)} + \epsilon_2 \right)$ satisfying the power constraints \eqref{EqPowerConstraints} with $i \in \lbrace a, b \rbrace$ and for all $s \in \mathcal{S}$, $\epsilon_s > 0$.
Then, without any loss of generality, for all $i \in \lbrace a, b \rbrace$ and for all $k \in \mathcal{K}$, we can write the sets $\mathcal{P}^{(i)}_k$ as follows,
\begin{eqnarray}
\label{EqStrategySetGa}    \mathcal{P}_k^{(a)} &=& \left\lbrace \left(p_{k}^{(1)},p_{k}^{(2)}\right) \in \mathds{R}^{2}: p_{k}^{(1)} = \alpha_k p_{k,\max} \text{ and } \right.\\
\nonumber
& & \left.p_{k}^{(2)} = p_{k,\max}(1 - \alpha_k), \text{ with } \alpha_k \in \left[0,1\right] \right\rbrace,\\
\label{EqStrategySetGb} \mathcal{P}_k^{(b)} &=& \left\lbrace \left(p_{k}^{(1)},p_{k}^{(2)}\right) \in \mathds{R}^{2}: p_{k}^{(1)} = \alpha_k p_{k,\max} \text{ and } \right.\\
\nonumber
& & \left.p_{k}^{(2)} = p_{k,\max}(1 - \alpha_k), \text{ with } \alpha_k \in \left\lbrace 0,1\right\rbrace \right\rbrace.
\end{eqnarray}

In the following section, both Def. \ref{DefNE} and Def. \ref{DefBR} are used to study the set of NE of both games $\mathcal{G}^{(i)}$, with $i \in \lbrace a, b \rbrace$.

\section{The Power Allocation Game}\label{SecGa}

In this section, we analyse the existence and the uniqueness of the NE in the PA game $\gameone$.

\subsection{Existence of the NE} 

The main result regarding the existence of the NE is stated in the following theorem.

\begin{theorem}[Existence of NE in the PA Game]\label{TheoremExistenceNEinGa}\emph{
The game $\gameone$ has always at least one NE in pure strategies.
}
\end{theorem}

\noindent
The proof of Theorem \ref{TheoremExistenceNEinGa} follows immediately from \emph{Theorem $1$ in \cite{Rosen-65}}. Note that, for all $k \in \mathcal{K}$, the utility function \eqref{EqUtility} of player $k$ is continuous and concave over the set of actions $\mathcal{P}^{(b)}_k$ and the set $\mathcal{P}^{(b)}$ is compact and convex. An alternative proof can be obtained from \emph{Theorem 3.2 in \cite{belmega-gamenets-2009}} or \emph{Theorem 1 in \cite{scutari-jsac-2008}}.

\subsection{Multiplicity of the NE}

In this subsection, we determine the number of possible NE which the game $\gameone$ can possess. In particular, it is found that, with probability one, it is possible to observe only one or three NE.  Finally, some sufficient conditions for observing a unique NE are stated.
\begin{theorem}[Multiplicity of NE]\label{TheoremMultiplicityNEinGa}\emph{
The game $\gameone$ might have either one, two, three or infinitely many NE in pure strategies. 
}
\end{theorem}

The proof of Theorem \ref{TheoremMultiplicityNEinGa} follows the same line of the proof of \emph{Theorem 3.3 in \cite{belmega-gamenets-2009}} and it is divided in two steps. First, we obtain an explicit expression for the BR correspondence (Def. \ref{DefBR}) for both players; second, the set of solutions of the fixed point equation in \eqref{EqFixPointNE} is determined.  
That is, for all $k\in\mathcal{K}$, let the action profile $\bs{p}_k = \left(p_{k}^{(1)},p_{k}^{(2)}\right)$ be written as follows $ \bs{p}_k= p_{k,\max} \left( \alpha_k,1 - \alpha_k\right)$, with $0 \leqslant \alpha_k \leqslant 1$, representing the fraction of power that transmitter $k$ uses over channel one. Then, with a slight abuse of notation, \eqref{EqUtility} can be written as follows, 
\begin{eqnarray}\label{u_alfa}
\nonumber
u_k(\alpha_k,\alpha_{-k}) & = & 
\log_2\left( 1+ \frac{\alpha_k g_{k,k}^{(1)}}{1+\alpha_{-k} g_{k,-k}^{(1)}}\right)+ \\
& &
\log_2\left( 1+ \frac{(1-\alpha_k) g_{k,k}^{(2)}}{1+(1-\alpha_{-k}) g_{k,-k}^{(2)}}\right),
\end{eqnarray}
where,
\begin{equation}
g_{j,k}^{(s)}=p_{k,max} \frac{\vert h_{j,k}\vert^{2}}{(\sigma_{j}^{(s)})^2}.
\end{equation}
Hence, the best response of player $k$ to the action $\bs{p}_{-k} \in \mathcal{P}_{-k}^{(a)}$, denoted by $p_{k,\max} \left(\alpha_k^{*},(1-\alpha_k^{*})\right)$, with $0 \leqslant \alpha_k^* \leqslant 1$, can be written in terms of $\alpha_{-k}$. For instance,
\begin{eqnarray}
\nonumber
	\BR^{(a)}_{k}(\bs{p}_{-k}) & = & \BR^{(a)}_{k}\left(p_{-k,\max} \left(\alpha_{-k},1-\alpha_{-k}\right)\right),\\
\label{Eq}
	& = & p_{k,\max} \left(\alpha_k^{*},(1-\alpha_k^{*})\right),
\end{eqnarray}
where $\alpha_k^*$ is the optimal fraction of transmit power transmitter $k$ must use over channel $1$. Following Def. \ref{DefBR}, we obtain
\begin{eqnarray}\label{EqAlphaStar}
	\alpha_k^{*} = \left\lbrace \begin{array}{lcl}
  	c_k\alpha_{-k}+d_k 	& \text{ if } & c_k\alpha_{-k}+d_{k}\in[0,1] \\
  	0  & \text{ if } 	& c_k\alpha_{-k}+d_{k} < 0 \\
  	1  & \text{ if } 	& c_k\alpha_{-k}+d_{k} > 1
	\end{array} \right.,
\end{eqnarray}
where,
\begin{equation}
c_k=-\frac{1}{2}\left(\frac{g_{k,-k}^{(1)}}{g_{k,k}^{(1)}}+\frac{g_{k,-k}^{(2)}}{g_{k,k}^{(2)}}\right)
\end{equation}
\begin{equation}
d_{k}=\frac{g_{k,k}^{(1)}(1+g_{k,-k}^{(2)})+g_{k,k}^{(2)}(g_{k,k}^{(1)}-1)}{2g_{k,k}^{(1)}g_{k,k}^{(2)}}.
\end{equation}
Once an explicit expression has been obtained for the BR correspondence of each player, the set of NE corresponds to the set of solutions to the fixed point inclusion \eqref{EqFixPointNE}.
\begin{figure}
\centering
\psfrag{A}{(a)}
\psfrag{B}{(b)}
\psfrag{C}{(c)}
\psfrag{D}{(d)}
\psfrag{E}{(e)}
\psfrag{Y}{$\alpha_1$}
\psfrag{X}{$\alpha_2$}
\psfrag{X1}{$1$}
\psfrag{Y1}{$1$}
\includegraphics[width=\linewidth]{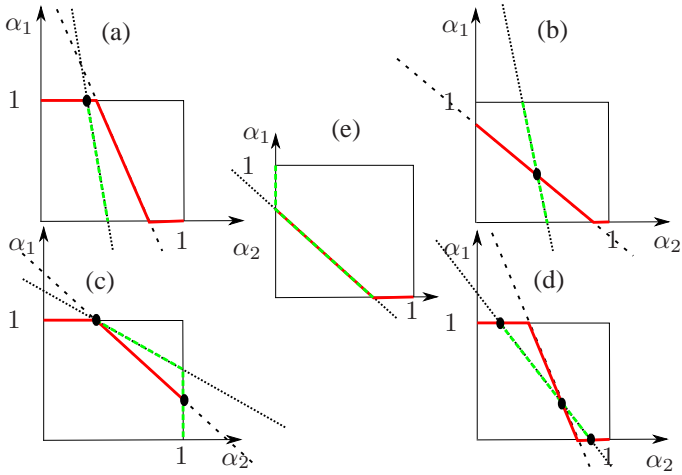}
\caption{The mappings $\alpha_k^*: [0,1] \rightarrow [0,1]$, with $k =1$ and $k = 2$ are represented by the red straight line and the dashed green line, respectively. The black dashed and dotted lines represent the lines in \eqref{gamma_def}, with $k = 1$ and $k=2$, respectively. Fig. (a) and (b) represents the case of unique NE, Fig. (c) and (d) represent the case of $2$ and $3$ NE, respectively. Fig. (e) represents the case of infinitely many NE.}
\label{FigBR}
\end{figure}
In Fig. \ref{FigBR}, we plot the mappings $\alpha_k^*: [0,1] \rightarrow [0,1]$, as defined by \eqref{EqAlphaStar}. Therein, any crossing point of both graphs is a solution of \eqref{EqFixPointNE}. Note that from Fig. \ref{FigBR}, it becomes evident that the number of intersection points can be either $1$, $2$, $3$ or infinitely many. 

\noindent
For a further analysis, we denote by $\bs{\alpha}^{\dagger}=(\alpha_{1}^{\dagger}$,$\alpha_{2}^{\dagger})$ the intersection point of the two lines,
\begin{equation}\label{gamma_def}
\gamma_{k} : \alpha_{1}=m_k\alpha_{2} + q_k;
\end{equation}
with $k \in \left\lbrace1,2\right\rbrace$, where $m_1=c_1$, $q_1=d_1$, $m_2=\frac{1}{c_{2}}$ and $q_2=-\frac{d_{2}}{c_{2}}$.
Hence, it follows that,
\begin{equation}\label{intersection point}
\begin{array}{lcl}
\alpha_{2}^{\dagger} & = & \frac{q_2-q_1}{m_1-m_2}=\frac{d_1c_2+d_2}{1-c_1c_2},\\
\alpha_{1}^{\dagger} & = & m_1\alpha_{2}^{\dagger}+q_1=\frac{d_2c_1+d_1}{1-c_1c_2}.
\end{array}
\end{equation}
A geometrical analysis of Fig. \ref{FigBR} leads to the following conclusions:\\
$(i)$ if $\bs{\alpha}^\dagger \not\in [0,1]^2$ then there is only one NE, where one player uses only one channel, while the other uses a water-filling PA vector \cite{Tse-05}. See Fig. \ref{FigBR} (a).\\
%
%
$(ii)$ if the following two conditions are met:
\begin{itemize}
\item $\exists k \in \mathcal{K}$, such that $\alpha^{\dagger}_k \in \lbrace 0,1 \rbrace$
and $\alpha^{\dagger}_{-k} \in [0,1]$
\item $\vert m_1 \vert > \vert m_2 \vert$
\end{itemize}
then the game has two NE. See Fig. \ref{FigBR} (c).\\
$(iii)$ if the two lines, $\gamma_1$ and $\gamma_2$, overlap, then the system has infinitely many NE. See Fig. \ref{FigBR} (e).\\
$(iv)$ if $\bs{\alpha}^{\dagger} \in [0,1]^2$ and $\frac{m_1}{m_2} > 1$ then there are three NE. See Fig. \ref{FigBR} (d)\\
$(v)$ if $\bs{\alpha}^{\dagger} \in [0,1]^2$ and $\frac{m_1}{m_2} < 1$ then there is only one NE. See Fig. \ref{FigBR} (b).
\\\noindent
Note that the next corollary follows immediately from conclusions $(ii)$ and $(iii)$, since the channels are random variables drawn from continuous distributions, and thus, the corresponding conditions are zero probability events.
\begin{corollary}\label{CorolMultiplicityNEinGa}\emph{
The game $\gameone$ has, with probability one, either one or three NE in pure strategies. 
}
\end{corollary}

\noindent
In the following, we provide sufficient conditions to observe a unique NE.

\begin{theorem}[Uniqueness of NE in the PA Game]\label{TheoremUniquenessNEinGa}\emph{
The game $\gameone$ has one NE if and only if at least one of the following conditions is satisfied.
\begin{eqnarray}
(a) & & (\rho_{1}^{(1)}+\rho_{1}^{(2)})(\rho_{2}^{(1)}+\rho_{2}^{(2)})<4\label{uniqcond}\\
(b)& & \exists k \in \mathcal{K}: \; \alpha_{k}^{\dagger}<0 \text{ or } \alpha_{k}^{\dagger}>1\label{uniqcond2}
\end{eqnarray}
where
\begin{equation}\label{rho_def}
\forall (k,s) \in \mathcal{K}\times\mathcal{S}, \rho_{k}^{(s)} = \frac{\vert h_{k,-k}^{(s)}\vert^2}{\vert h_{k,k}^{(s)}\vert^2}.
\end{equation}
}
\end{theorem}
The proof of Theorem \ref{TheoremUniquenessNEinGa} is as follows. From conditions $(i)$ - $(iv)$, a sufficient condition to observe a unique NE can be implied: 
\begin{equation}\label{EqUniquenessCondition}
\vert m_{2}\vert > \vert m_{1}\vert.
\end{equation} 
Then, following equation \eqref{rho_def}, it is possible to write
\begin{equation}\label{m1def}
m_{1}=-\frac{p_{2,max}}{2p_{1,max}}(\rho_{1}^{(2)}+\rho_{1}^{(1)}),
\end{equation}
and
\begin{equation}\label{m2def}
m_{2}=\frac{-2p_{2,max}}{p_{1,max}(\rho_{2}^{(2)}+\rho_{2}^{(1)})},
\end{equation}
and thus, replacing \eqref{m1def} and \eqref{m2def} in \eqref{EqUniquenessCondition} yields equation \eqref{uniqcond}.
Condition $(b)$ is inferred by graphical arguments. The direct implication comes from (a) in Fig. \ref{FigBR}, for the reverse one we have to notice that, with probability one, we can observe either one NE ((a) and (b) in Fig. \ref{FigBR}) or three ((d) in Fig. \ref{FigBR}). As a consequence, the uniqueness of the NE implies \eqref{uniqcond} or \eqref{uniqcond2}. 

\noindent
Note that, \eqref{uniqcond} represents the geometric average of the algebraic average of the ratios between the interfering and direct channels. Interestingly, it shows also that, if the direct channels are always stronger than the interfering ones, or if one transmitter-receiver couple is isolated from the other (i.e. $\exists (k,s)\in\mathcal{K}\times\mathcal{S}: g_{k,-k}^{(s)}=0$), then the NE is unique. 
Finally, we would like to point out the fact that condition \eqref{uniqcond} is in accordance both with the one in \cite{scutari-jsac-2008} and in \cite{jorswiek-gamenet-2009} for obseving a unique NE. However, the condition in \cite{scutari-jsac-2008} appears to be more restrictive, while the condition in \cite{jorswiek-gamenet-2009} can be easily deducted from \eqref{uniqcond}, by setting $\rho_1^{(1)}=0$ and $\rho_2^{(1)}=0$. Moreover, \eqref{uniqcond} insures the convergence of the BRD

\section{Channel Selection Game}\label{SecGb}

In this section, we study the channel selection game $\gametwo$. In this case, contrary to the PA the action space is a discrete set, thus the existence of a pure NE is not deducible from the application of \emph{Theorem $1$ in \cite{Rosen-65}}.
\noindent
Let the channel selection vector of player $k$, be denoted by $\bs{p}_k = p_{k,\max}\left(\alpha_k,1-\alpha_k\right)$, with $\alpha_k \in \lbrace 0, 1\rbrace$. In the following we will indifferently refer to the Channel Selection NE $\bs{p^*}=(p_1^*,p_2^*)$ as $\bs{\alpha^*}=(\alpha_1^*,\alpha_2^*)$ with $\alpha_k^*\in\left\lbrace 0,1\right\rbrace$.

Hence, all the outcomes of the game can be described by the table hereunder:
\begin{figure}[h]
\begin{center}
$\begin{array}{|c|c|c|}\hline
\scriptstyle Tx_1 \backslash Tx_2 &  \alpha_2 = 1 & \alpha_2 = 0 \\ \hline
\alpha_1 = 1 &  \left(u_1(1,1),u_2(1,1)\right) & \left(u_1(1,0),u_2(0,1)\right)\\ \hline
\alpha_1 = 0 &  \left(u_1(0,1),u_2(1,0)\right) & \left(u_1(0,0),u_2(0,0)\right)\\ \hline
\end{array}$
\end{center}
\caption{Utility obtained by player $1$ and $2$, where $u_k$ is defined by \eqref{u_alfa}. Player $1$ chooses rows and player $2$ chooses columns.}
\label{TabNE}
\end{figure}\\
In the following, we study the existence and the multiplicity of the NE of the game $\gametwo$.

\subsection{Existence of the NE}
The main result regarding the existence of the game $\gametwo$ is the following.
\begin{theorem}[Existence of NE in the CS Game]\label{TheoremExistenceNEinGb}\emph{
The game $\gametwo$ has always at least one NE in pure strategies.
}
\end{theorem}
The proof of Theorem \ref{TheoremExistenceNEinGb} follows from showing that for any vector $\bs{g} = \left(g_{j,k}^{(s)} \right)_{(j,k,s)\in\mathcal{K}^2\times\mathcal{S}} \in \mathds{R}^8$, there always exists an outcome $\bs{\alpha}^* = (\alpha_1^*,\alpha_2^*)$, which is a NE. To prove it we will perform an exhaustive search.
\noindent
Without any loss of generality, let us assume $g_{1,1}^{(1)} > g_{1,1}^{(2)}$ then,
\begin{itemize}
\item if $g_{2,2}^{(2)}>g_{2,2}^{(1)}$, then $\bs{\alpha^{*}}=(1,0)$ is a NE;
\item if $g_{2,2}^{(2)}<g_{2,2}^{(1)}$, then
\begin{itemize}
\item if $g_{2,2}^{(1)}<g_{2,2}^{(2)}(1+g_{2,1}^{(1)})$, then $\bs{\alpha^{*}}=(1,0)$ is a NE;
\item if $g_{2,2}^{(1)}>g_{2,2}^{(2)}(1+g_{2,1}^{(1)})$, then
\begin{itemize}
\item $g_{1,1}^{(1)}>g_{1,1}^{(2)}(1+g_{1,2}^{(1)})$, then $\bs{\alpha^{*}}=(1,1)$ is a NE;
\item $g_{1,1}^{(1)}<g_{1,1}^{(2)}(1+g_{1,2}^{(1)})$, then $\bs{\alpha^{*}}=(1,0)$ is a NE;
\end{itemize}
\end{itemize}
\end{itemize}
We analyse all the possible NE in the following subsection.
\subsection{Multiplicity of the NE}
In the game $\gametwo$, depending on the channel realizations, any of the four outcomes of the game can be a NE as shown in Theorem \ref{NEinGb}. However, as we shall see, the game may have either one or two NE. 

\begin{theorem}[NE in $\gametwo$]\label{NEinGb}\emph{Consider the game $\gametwo$ and let $\bs{\alpha^{*}} = (\alpha_{1}^{*},\alpha_{2}^{*})$ identify the channel selection $p_{k}^{*}=p_{k,max}(\alpha_{k}^{*},1-\alpha_{k}^{*})\forall k \in \mathcal{K}$.Then
\begin{enumerate}
\item $\bs{\alpha^{*}}=(1,1)$ is a NE if and only if
\begin{equation}\label{DS1}
\left\lbrace \begin{array}{lcl}
g_{1,1}^{(1)}&>&g_{1,1}^{(2)}(1+g_{1,2}^{(1)})\\
g_{2,2}^{(1)}&>&g_{2,2}^{(2)}(1+g_{2,1}^{(1)})
\end{array}\right.
\end{equation}
\item $\bs{\alpha^{*}}=(0,0)$ is a NE if and only if
\begin{equation}\label{DS2}
\left\lbrace \begin{array}{lcl}
g_{1,1}^{(2)}&>&g_{1,1}^{(1)}(1+g_{1,2}^{(2)})\\
g_{2,2}^{(2)}&>&g_{2,2}^{(1)}(1+g_{2,1}^{(2)})
\end{array}\right.
\end{equation}
\item $\bs{\alpha^{*}}=(0,1)$ is a NE if and only if
\begin{equation}
\left\lbrace \begin{array}{lcl}
g_{1,1}^{(2)}(1+g_{1,2}^{(1)})&>&g_{1,1}^{(1)}\\
g_{2,2}^{(1)}(1+g_{2,1}^{(2)})&>&g_{2,2}^{(2)}
\end{array}\right.
\end{equation}
\item $\bs{\alpha^{*}}=(1,0)$ is a NE if and only if
\begin{equation}
\left\lbrace \begin{array}{lcl}
g_{1,1}^{(1)}(1+g_{1,2}^{(2)})&>&g_{1,1}^{(2)}\\
g_{2,2}^{(2)}(1+g_{2,1}^{(1)})&>&g_{2,2}^{(1)}
\end{array}\right.
\end{equation}
\end{enumerate}
}
\end{theorem}
The proof of Theorem \ref{NEinGb} is an immediate result from Fig. \ref{TabNE} and Def. \ref{DefNE}.
An important conclusion which follows immediately from Theorem \ref{NEinGb} is the following.
\begin{corollary}[Multiplicity of the NE in the CS game]\label{CS_anticoordination}\emph{The game $\gametwo$ has always either one or two NE in pure strategies.
}
\end{corollary}
This result follows from the fact that if there exists a player with a dominant strategy, the game $\gametwo$ has a unique NE. If none of the players possesses a dominant strategy, the game $\gametwo$ is an anti-coordination game with two NE $\bs{\alpha^{*}}=(1,0)$ and $\bs{\alpha^{*}}=(0,1)$.
\section{Numerical Results}\label{SecSimulations}
The purpose of this section is two-fold. First, we provide numerical approximations of the probability of observing either one or three NE in the game $\gameone$, and the probability of observing one or two NE in $\gametwo$. Second, we provide numerical calculations of the utilities achieved in the games $\gameone$ and $\gametwo$, in order to evaluate which game brings the highest system spectral efficiency at the equilibrium, i.e., the sum of all individual spectral efficiencies \eqref{EqUtility}. 

\begin{figure}
\centering
\includegraphics[width=0.5\textwidth]{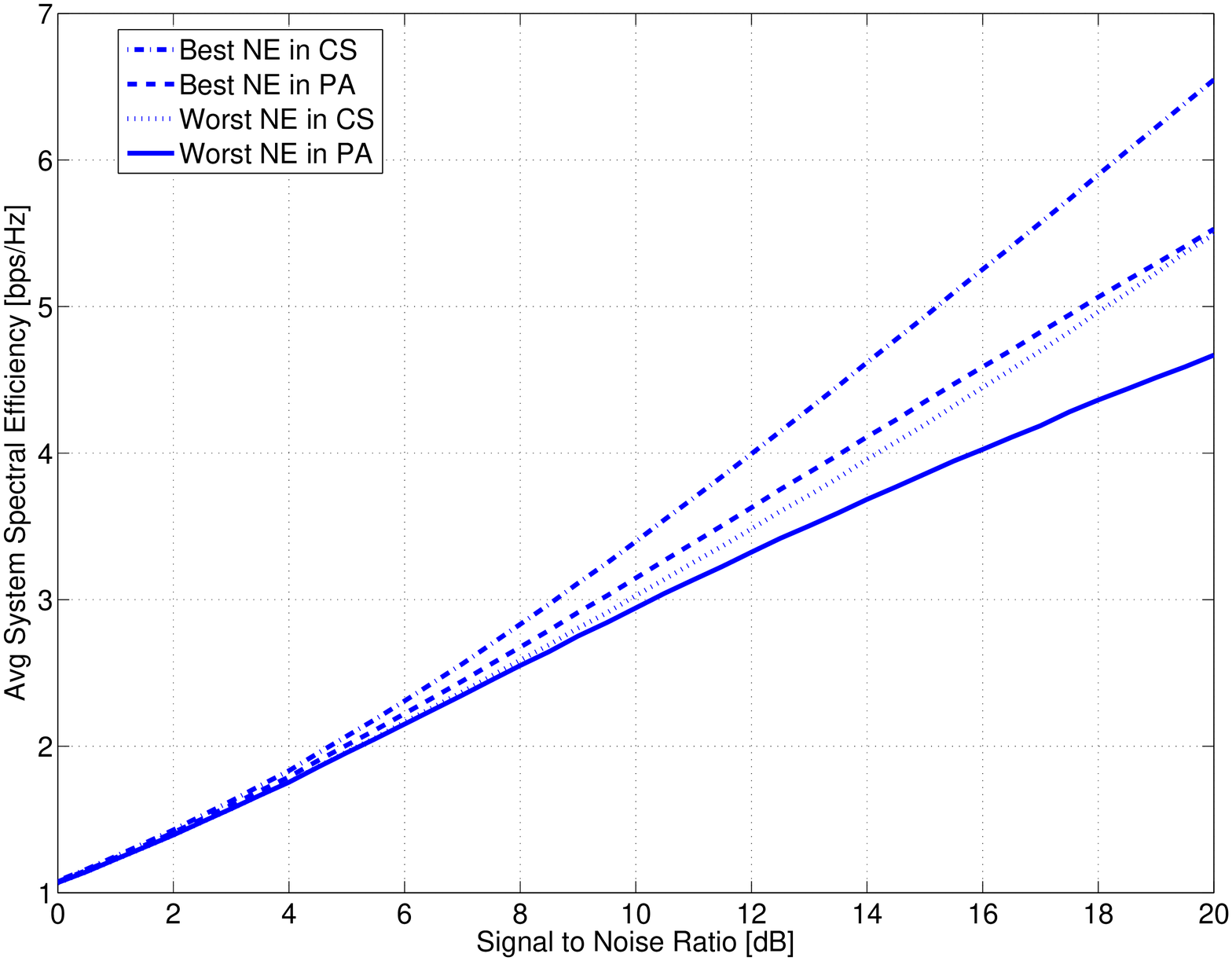} 
\caption{Average sum-utility (system spectral efficiency) of the best and worst NE of the games $\gameone$ and $\gametwo$.}
\label{avg_def}
\end{figure}

\begin{figure}
\centering
\includegraphics[width=0.5\textwidth]{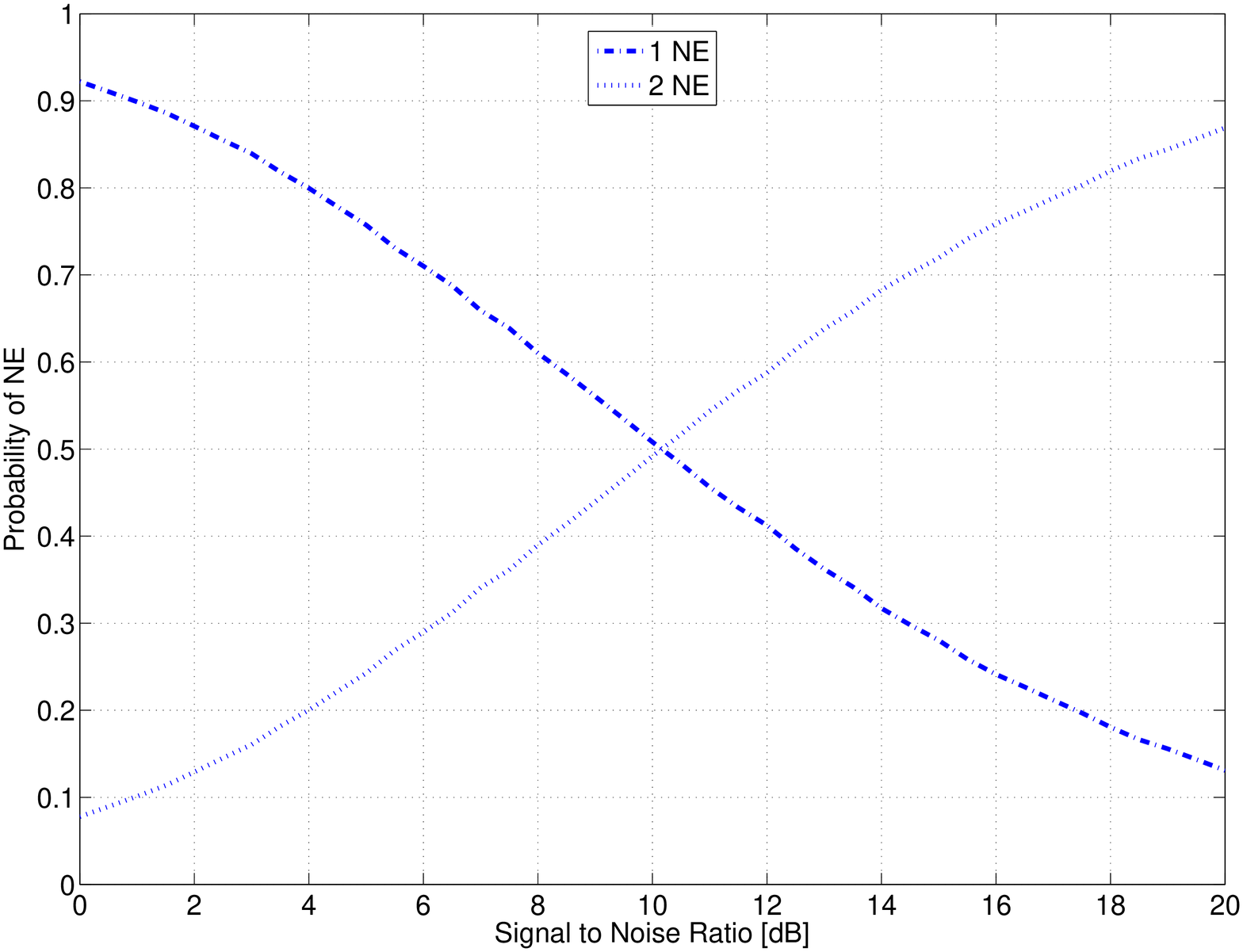} 
\caption{Probability of observing either one or two NE in the game $\gametwo$.}
\label{CS_prob}
\end{figure}

\noindent
In the first experiment, we generate $10^6$ vectors of channel realizations $\bs{g} = \left(g_{j,k}^{(s)} \right)_{(j,k,s)\in\mathcal{K}^2\times\mathcal{S}} \in \mathds{R}^8$ and, for each realization, the number of NE of the corresponding game is calculated. In $\gameone$, when only one NE is observed, we distinguish between the case where $\exists k\in \mathcal{K}: \alpha_k^* \in \lbrace 0, 1 \rbrace$ (see Fig. \ref{FigBR} (a)) and the case where $\bs{\alpha}^* \in \left[0,1\right]^2$ (See Fig. \ref{FigBR} (b)). The results of this Monte Carlo simulation are reported in Fig. \ref{avg_def}.
About $\gametwo$, the results are reported in Fig. \ref{CS_prob}. As for $\gameone$, the probability of observing a multiple NE increases with the SNR. This is easily explicable by noting that low SNR also means that the interference is negligible when compared to the the noise and vice-versa. As a consequence, when the noise is the major concern (low SNR regime) the transmitter will try to optimize its spectral efficiency by selecting the least noisy channel regardless of the interference. When, on the contrary, the major concern is the interference (high SNR regime) then avoiding the channel used by the other transmitter becomes the priority. 
Note that the game $\gameone$ has only one NE with a high probability, however, three NE can be observed with a non-negligible one. This result implies that designing algorithms for achieving NE (see \cite{Scutari-Algorithms-2008}) in decentralized networks requires to tackle the problem of equilibrium selection, which is a problem that has been neglected in most recent literature \cite{Scutari-Algorithms-2008, Scutari-Palomar-09}.
\begin{figure}
\centering
\includegraphics[width=0.5\textwidth]{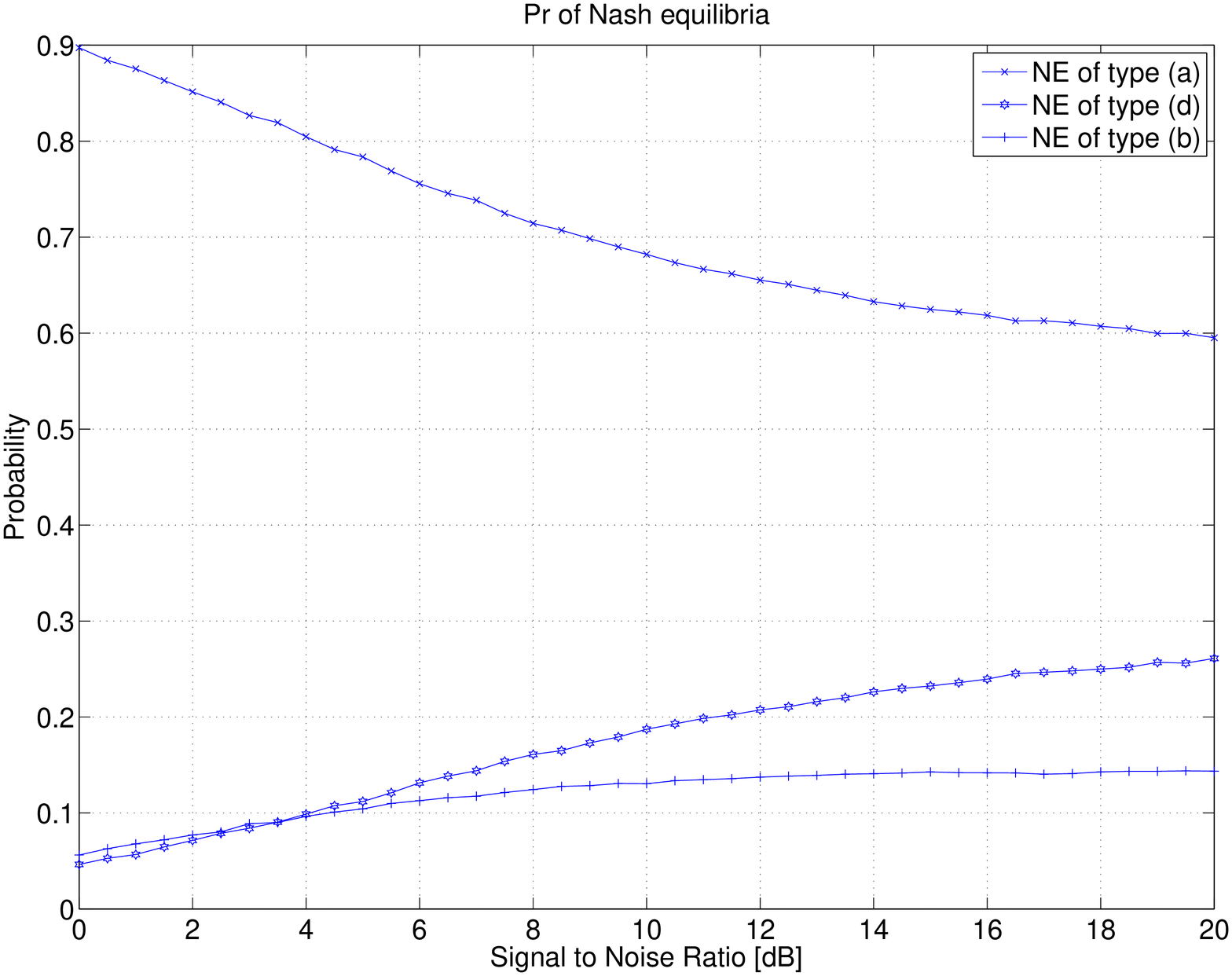} 
\caption{Probability of observing either one or three NE in the game $\gameone$. We refer to type (a), the case where there exists a unique NE such that $\exists k\in \mathcal{K}: \alpha_k^* \in \lbrace 0, 1 \rbrace$ (see Fig \ref{FigBR} (a)). Type (b) refers to the case where there exists a unique NE and $\forall k \in \mathcal{K}$, $\bs{\alpha}^* \in \left[0,1\right]^2$ (See Fig \ref{FigBR} (b)). Type (d) refers to the case where there exists three NE.}
\label{CSvsPA}
\end{figure}

In the second experiment, we generate $10^6$ vectors of channel realizations $\bs{g} = \left(g_{j,k}^{(s)} \right)_{(j,k,s)\in\mathcal{K}^2\times\mathcal{S}} \in \mathds{R}^8$ and for each realization, the utility achieved in each of the NE of the games $\gameone$ and $\gametwo$ is calculated. In Fig. \ref{avg_def}, we report the average sum-utility achieved in any of the NE of the games $\gameone$ and $\gametwo$. Note that a higher spectral efficiency is observed when transmitters are limited to use only one channel. This result can be interpreted as a Braess paradox \cite{Braess-69} and generalizes the founds in \cite{Perlaza-Gamecomm-09}, \cite{Altman-wiopt-2008} and \cite {Altman-wiopt-08b}. Here, reducing the set of actions of each player leads to a higher sum-utility, i.e., a higher system spectral efficiency.  


\section{conclusions}\label{SecConclusions}
In this paper, we presented a game theoretical analysis of the $2$-dimensional parallel interference channel with two transmitter-receiver pairs. Two games were analysed. First, the power allocation game where transmitters can simultaneously use both channels. Second, the channel selection game, where transmitters use only one of the available channels at a time. Here, the number of NE in the PA game has been proved to be either $1$ or $3$ depending on the exact channel realizations. \balance However, it has been also shown that with zero-probability, it is possible to observe either $2$ or infinitely many NE. Regarding the CS game, depending on the channel realizations, any feasible channel selection might be a NE. Here, we provide conditions over the channel realizations for every case. In particular, the number of NE in pure strategies in the CS game is either one or two depending on the channel realizations. Finally, we showed, by using Monte-Carlo simulations, that the best and worst average system spectral efficiency achieved in equilibrium in the CS game is better than the best and worst average system spectral efficiency achieved in equilibrium in the PA game. 

\section{ACKNOWLEDGMENT}
This work is supported by the European Commission in the framework of the FP7 Network of Excellence in Wireless Communications NEWCOM++.

\bibliographystyle{IEEEtran}
\bibliography{GT}
\end{document}